\documentclass[useAMS,usenatbib]{mn2e} 
 \usepackage{graphicx}

 \title[Spiral Structure of the Milky Way]
 {The Milky Way Spiral Structure Parameters from Data on Masers and Selected Open Clusters}
 \author[V. V. Bobylev and A. T. Bajkova]{V. V. Bobylev$^{1,2}$
 \thanks{E-mail: vbobylev@gao.spb.ru} and A. T. Bajkova$^{1}$\\
 $^{1}$Central (Pulkovo) Astronomical Observatory of RAS, 65/1 Pulkovskoye Chaussee, Saint Petersburg, 196140, Russia\\
 $^{2}$ Sobolev Astronomical Institute, St. Petersburg State University, Bibliotechnaya pl. 2, St. Petersburg, 198504, Russia}

\begin{document}
 \date{Accepted 2013 MONTH XX.} 
 \pagerange{\pageref{firstpage}--\pageref{lastpage}} \pubyear{2013}
 \maketitle
 \label{firstpage}

\begin{abstract}
To estimate the parameters of the Galactic spiral structure~--
namely, the pitch angle $i$ and the number of spiral arms,~-- data
on Galactic masers with known trigonometric parallaxes were used.
We applied the well-known method based on analysis of the
``position angle -- distance logarithm'' diagram. Estimates of the
pitch angle $i$ obtained from four segments of different arms
belonging to the global Galactic structure are self-consistent and
close to $i=-13^\circ\pm1^\circ$. The segment which is most
interesting is that of the Outer arm. It contains only three
masers. Hence, in order to obtain correct estimates, we also used
the data on 12 very young star clusters with distances determined
by Camargo et al. from infrared photometry. The estimates obtained
allow us to conclude in favor of the four-armed model of the
Galactic spiral structure.
\end{abstract}


\section{Introduction}
Up to now, the question of the number of spiral arms in our Galaxy
still has no unequivocal answer. As follows from the analysis of
spatial distribution of young Galactic objects (young stars,
star-forming regions, open star clusters, hydrogen clouds), two-,
three- and four-arm models of the Galactic spiral structure are
possible~\citep{Russeil03,Valee08,Hou09,Efremov11,Francis12}. More
complicated models are also known~-- for example, the kinematical
model of~\cite{Lepine01}, where two- and four-arm patterns are
combined in the Solar neighborhood. According to
\cite{Englmaier11}, the distribution of neutral hydrogen suggests
that there is a two-arm pattern in the inner part ($R<R_0$) of the
Milky Way which splits into four arms in its outer part ($R>R_0$).
Note also the spiral ring model~\cite{Melnik09} that contains two
outer rings stretched in the perpendicular and parallel directions
to the central bar, the inner ring stretched parallel to the bar,
and also two small spiral arm fragments.

In our previous kinematic analysis aimed at determining
the spiral density wave parameters, we adopted the most simple
two-arm model~\citep{b08,bb10,bb11}, which allowed us to estimate
the spiral pitch angle $i$ directly from the estimate of the
spiral wavelength $\lambda$~\citep{bb12} that binds both
parameters~-- pitch angle and the number of spiral arms. Therefore, a
direct method of estimating the pitch angle would be of great
interest.

In this paper, a study of the Galactic spiral pattern is done on
the basis of data on spatial distribution of the youngest Galactic
objects tracing the Galactic spiral arms. The data consist of
coordinates and trigonometric parallaxes of Galactic masers
obtained by several VLBI groups during long-term radio
interferometric campaigns within the various
projects~\citep{Reid09a,BesseL11,Honma12}.

In addition, we use 12 very young star clusters recently
discovered in the Outer spiral arm of the Galaxy, which distances
and ages were estimated with high accuracy from infrared
photometry~\citep{Camargo13}. This allowed us to considerably
extend the sample of objects in the Outer arm for determining the
pitch angle.

Thus, the aim of this work was to determine Galactic spiral
structure parameters from young objects (basically masers as well
as young clusters in the Outer arm) distributed in a wide range of
galactocentric distances and position angles. The use of the
``position angle~-- distance logarithm'' diagram allowed us to
directly estimate the spiral arm pitch angle and determine the
number of spiral arms.

\section[]{Method}\label{Techniq}
The equation describing the position of a Galactic object on the
logarithmic spiral can be written in the following way:
 \begin{equation}
 R=a_0 e^{(\theta-\theta_0)\tan i},
 \label{spiral-1}
 \end{equation}
where $a_0>0$, $\theta$~is the object's position angle measured in
the direction of Galactic rotation: $\tan\theta = y/(R_0-x)$,
where $x,y$~are Galactic heliocentric rectangular coordinates of
the object; $\theta_0$ is angle at which $R=a_0$; $i$ is pitch
angle ($i<0$ for leading spirals) which is related to other spiral
structure parameters as:
\begin{equation}
  \tan i=\frac{m\lambda}{2\pi R_0},
 \label{a-04}
\end{equation}
where $m$~is the number of spiral arms, $\lambda$~is the wavelength of
spiral wave which is equal to the distance (in galactocentric radial
direction) between adjacent segments of spiral arms in the solar
neighborhood, $R_0$~is galactocentric distance of the Sun which is
adopted to be $R_0=8$~kpc. Radial phase of the spiral wave
$\chi$ is
 \begin{equation}
   \chi=m[\cot (i)\ln (R/R_0)-(\theta-\theta_0)]+\chi_\odot,
 \label{chi-creze}
 \end{equation}
where   $\chi_\odot$~is  radial phase of the Sun in the spiral
wave.

Putting $a_0=R_0$ in Equation~(\ref{spiral-1}), we can estimate
the value of pitch angle $i$ as
\begin{equation}
  \tan i=\frac{\ln (R/R_0)}{\theta-\theta_0},
 \label{spiral-04}
\end{equation}
where, obviously, $\theta_0=0^\circ$. For this purpose, a
``position angle~-- distance logarithm'' diagram is constructed, where
arms of a logarithmic spiral are presented as line segments.
Such method is widely used for studying the
Galactic spiral structure based on the various object
data~\citep{Popova05,Xu13}. The advantage of this approach is
that the estimate of pitch angle $i$ does not depend on the number
of spiral arms.

\section[]{Data}\label{Data}
We use coordinates and trigonometric parallaxes of masers measured
by VLBI with errors of less than 10\% in average. These
masers are connected with very young objects (basically proto
stars of high masses, but there are ones with low masses too; a
number of massive super giants are known as well) located in active
star-forming regions.

One of such observational campaigns is the Japanese project VERA
(VLBI Exploration of Radio Astrometry) for observations of water
(H$_2$O) Galactic masers at 22~GHz~\citep{Hirota07} and SiO masers
(such masers are very rare among young objects) at
43~GHz~\citep{Kim08}.

Water and methanol (CH$_3$OH) maser parallaxes are observed in USA
(VLBA) at 22~GHz and 12~GHz~\citep{Reid09a}. Methanol masers are
observed also in the framework of the European VLBI
network~\citep{Rygl10}. Both these projects are joined together in
the BeSSeL program~\citep{BesseL11}.

VLBI observations of radio stars in continuum at
8.4~GHz~\citep{Dzib11} are carried out with the same goals.

\begin{figure*}
  \includegraphics[width=1.90\columnwidth]{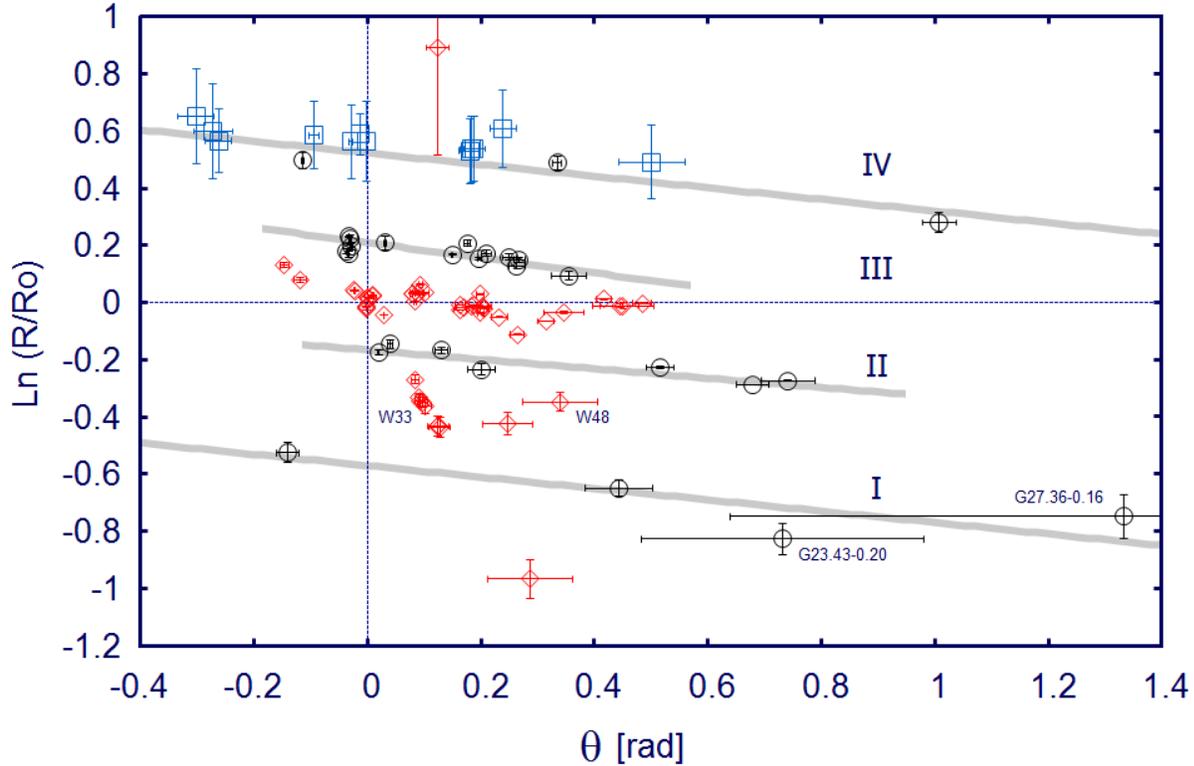}
   \caption{Distance logarithm $\ln(R/R_0)$
versus position angle $\theta$, location of the Sun is indicated
by a dotted line, spiral arms are marked by Roman numerals
(I--IV), masers are indicated by circles, young open star clusters
in the Outer spiral arm (IV) are shown by gray squares, masers
that were not used to define the characteristics of spiral arms
are indicated by  diamonds. }
  \label{f1}
 \end{figure*}

Complete information on 54 masers with measured trigonometric
parallaxes is given in the papers by~\cite{bb12} and
\cite{Stepanishchev13}. Apart from that, we use data on new sources as well
as data from the recent measurements published in the following
papers:

\begin{enumerate}
\item a study by~\cite{Wu12} of the star-forming region
RCW122 (G348.70$-$1.04). It is a very important source that
considerably widens the available range of position angles, which is very
important for estimation of the pitch angle value (in Fig.~\ref{f1}, it
belongs to arm~I and has the following coordinates:
$\ln(R/R_0)=-0.52$ and $\theta=-0.14$~rad.);

\item \cite{Imai12} on the IRAS~22480$+$6002 source which is associated with
a massive super-giant of spectral class K from Perseus arm;

\item \cite{Sakai12} on the IRAS~05168$+$3634 source from Perseus arm;

\item \cite{Xu13} devoted to the study of peculiarities of the local spiral arm (Orion arm)
using data on 30 masers;

\item \cite{Immer13} containing parallax measurements for a number of
masers in star-forming regions W33 and G012.88$+$048.
\end{enumerate}

As a result, a sample of 82 sources has been compiled. The sample
was extended by the data on 12 very young star clusters which
distances and ages (2 Myr average, 10 Myr maximum) were estimated
by~\cite{Camargo13} with high accuracy from 2MASS infrared
photometry~\citep{Skrutskie06}. These objects are known as
nebula-embedded clusters. They have not been discovered in optical
until recently due to a strong absorption in their direction.
Actually, they are just young star associations and clusters with
gas still not having swept by supernovae explosions because
massive stars had no time to evolve to that stage. As it was shown
by~\cite{Camargo13}, all these objects along with large number of
massive stars of high luminosity contain many stars of T~Tau type,
what is typical to very young star clusters and associations. All
these clusters CBB10, CBB11, CBB12, CBB14, CBB15, FSR486, FSR831,
FSR843, FSR851, FSR909, FSR1099, and NGC1624 are located in the
Outer spiral arm.

\section[]{Results and Discussion}\label{Resalts}
\subsection[]{Grand-design structure}\label{grand}
In Fig.~\ref{f1}, the ``position angle~-- distance logarithm''
diagram is shown. When plotting Fig.~\ref{f1}, we visually
indicated the four possible arms. We considered the following
masers as belonging to arm I:
               G23.43$-$0.20, G23.01$-$0.41, G27.36$-$0.16 and RCW~122;
 those belonging to arm II: G14.33$-$0.64, G35.20$-$0.74, IRAS~19213$+$1723, W51,
               G5.89$-$0.39, G48.61$+$0.02 and MSXDC~G034.43$+$0.24;
 those belonging to arm III: IRAS~00420$+$5530, NGC~281-W, S~Per, W3-OH, S252A
               IRAS~06058$+$2138, IRAS~06061$+$2151, S255, AFGL~2789, NGC~7538,
               G192.16$-$3.84, IRAS~5168$+$3634, PZ~Cas and IRAS~22480$+$6002;
 and those belonging to arm  IV: WB89$-$437, S269~(G196.45$-$1.68) and G75.30$+$1.32.
Two masers marked in Fig.~\ref{f1} (G09.62$+$0.20 and
IRAS~05137$+$3919) that were located at the edges of the diagram
were not used. A group of masers associated with star-forming
regions W33 and W48 was not used as well since it is falling in
the interarm space (between arms I and II). All masers (37
sources) falling into the interarm space between arms II and III
(Fig.~\ref{f1}) belong to the local (Orion) arm. As for other
masers, there were no doubt in attributing them to one or another
arm.

The spiral arms characteristics were obtained from linear regression
$\ln(R/R_0)=a\cdot\theta+b$ (see eq, (\ref{spiral-04})). The
problem was solved both with unit weights (Table~\ref{t1}) and with
weights inversely proportional to the squared errors of
distances (Table~\ref{t2}).

The errors of galactocentric distance $\sigma_R$ and position
angle $\sigma_\theta$ can be easily found from the error of
heliocentric distance $\sigma_r$. Then the error of $\ln(R/R_0)$
is $\sigma_{\ln(R/R_0)}=\ln(1+\sigma_R/R_0)$.

When solving the problem of linear regression, the errors of
unknowns $a$ and $b$ were determined by the Monte Carlo method using
1000 random samples along both coordinates
$\ln(R/R_0)$ and $\theta$ varying within errors.

Weights for equations (we used weights equal to
$1/\sigma^2_{\ln(R/R_0)}$) are usually used in case of
heterogeneous data. As it has been found out, the use of weights
is especially relevant for arm IV where the data are really
mixed. As it is seen by comparing Table~\ref{t1} and
Table~\ref{t2}, this approach had a favorable effect on the
results obtained for arm IV because the error of $i$ has
decreased.

Parameters of the fit lines shown in Fig.~\ref{f1} were taken from
Table~\ref{t2}.

\begin{figure*}
   \includegraphics[width=1.40\columnwidth]{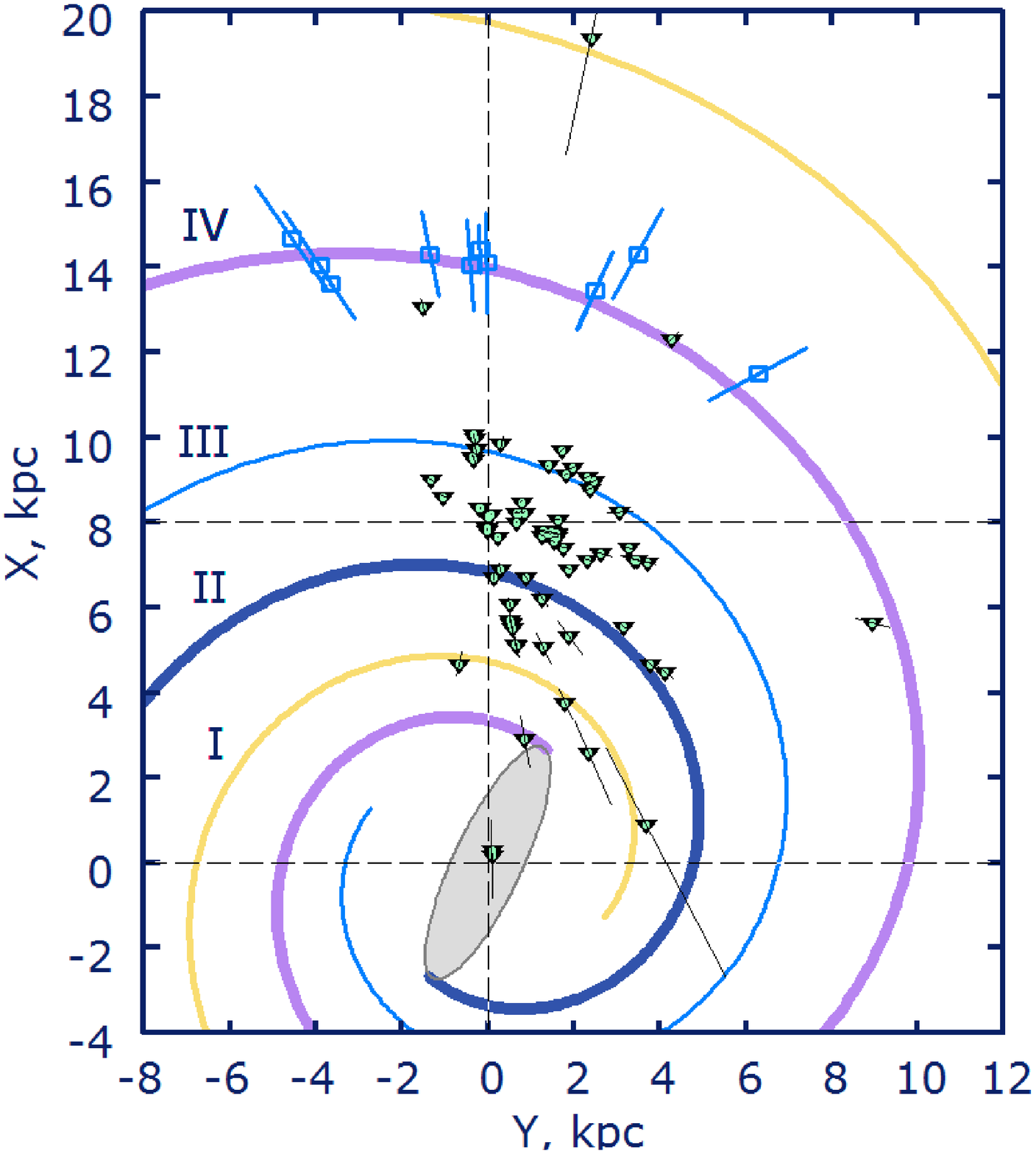}
      \caption{Four-arm spiral pattern of the Galaxy with pitch angle
$i=13^\circ$, location of the Sun and the center of the Galaxy are
indicated by dotted lines, masers are marked by black triangles,
young open star clusters in the Outer arm (IV) are shown by
squares, central bar is shown as an ellipse.}
         \label{f2}
\end{figure*}

In Fig.~\ref{f2}, a four-armed spiral pattern constructed for
pitch angle $i=-13^\circ$ is shown (see discussion below). Spiral
pattern can be easily drawn using (\ref{spiral-1}) for each $k$-th
order spiral arm with known pitch angle $i$ and parameters
$a_0=R_k, k=1,...,4$; each arm is drawn independently of the
others.

It is seen that the clusters from the paper by~\cite{Camargo13}
are good tracers of the Outer arm. We can see two masers from the
region Sgr~B2 in the center of the Galaxy~\citep{Reid09b}.

  \begin{table*}
   \begin{center}
    \caption{Parameters of linear regression $\ln(R/R_0)=a\cdot\theta+b$,
found using unit weights}
   \label{t1}
   {\small
   \begin{tabular}{|l|c|c|c|r|r|}      \hline
                Arm & $n_\star$ &     $a$          &    $b$           & $  i,$ deg    & $R_k$, kpc\\\hline
 I   (Scutum--Crux)        &  4 & $-0.174\pm0.099$ & $-0.597\pm0.041$ & $- 9.8\pm5.4$ & $ 4.40\pm0.24$ \\
 II  (Carina--Sagittarius) &  7 & $-0.163\pm0.014$ & $-0.161\pm0.008$ & $- 9.3\pm0.8$ & $ 6.81\pm0.26$ \\
 III (Perseus)             & 14 & $-0.207\pm0.024$ & $+0.200\pm0.005$ & $-11.7\pm1.3$ & $ 9.77\pm0.36$ \\
 IV  (Outer, or Cygnus)    & 15 & $-0.214\pm0.072$ & $+0.563\pm0.033$ & $-12.2\pm3.9$ & $14.02\pm0.69$ \\
 \hline
 simple average     &&&& $-10.7\pm0.7$ & \\
 weighted average   &&&& $- 9.9\pm0.6$ & \\
 \hline
      \end{tabular}}
     \end{center}
{\small Note. $R_k$ is the value of $b,$ in kpc and a point of
intersection of the k-th spiral arm with the direction from the
Sun to the center of the Galaxy ($\theta=0^\circ$).}
   \end{table*}

  \begin{table*}
   \begin{center}
    \caption{Parameters of linear regression $\ln(R/R_0)=a\cdot\theta+b$,
found using weights $1/\sigma^2_{\ln(R/R_0)}$}
   \label{t2}
   {\small
   \begin{tabular}{|l|c|c|c|r|r|}      \hline
 Arm & $n_\star$ & $a$ & $b$ & $i,$ deg & $R_k$, kpc\\\hline
 I   (Scutum--Crux)        &  4 & $-0.199\pm0.075$ & $-0.570\pm0.030$ & $-11.2\pm4.0$ & $ 4.52\pm0.21$ \\
 II  (Carina--Sagittarius) &  7 & $-0.163\pm0.040$ & $-0.166\pm0.026$ & $- 9.3\pm2.2$ & $ 6.78\pm0.32$ \\
 III (Perseus)             & 14 & $-0.265\pm0.015$ & $+0.210\pm0.002$ & $-14.8\pm0.8$ & $ 9.87\pm0.37$ \\
 IV  (Outer, or Cygnus)    & 15 & $-0.203\pm0.036$ & $+0.524\pm0.017$ & $-11.5\pm1.9$ & $13.51\pm0.54$ \\
 \hline
 simple average     &&&& $-11.7\pm1.1$ & \\
 weighted average   &&&& $-13.7\pm1.1$ & \\
 \hline
      \end{tabular}}
     \end{center}
   \end{table*}

\cite{Xu13} obtained an estimate of the pitch angle for Orion arm
$i=-10.1^\circ\pm2.7^\circ$ from 30 masers. All these masers are
shown in Fig.~\ref{f1} and Fig.~\ref{f2}.
Evolutionary status of the Local arm is discussed below.

As it is seen from Fig.~\ref{f2}, almost every object used in
estimations of the pitch angle belongs to a certain arm. If we
assume that our model of arms is true, then it is possible to
improve the statistics for arm~I, where two masers G23.43$-$0.20
and G27.36$-$0.16 have very large errors $\sigma_\theta$. To
achieve this, we can complement four masers located in the inner
part of the Galaxy with the most distant maser IRAS~05137$+$3919,
also belonging to this arm but located in the outer part of the
Galaxy. Then the position angle of this object is
$\theta=\theta-2\pi$, which widens dynamical range of position
angles to as large as $440^\circ$. As the final result, a new
estimate of the pitch angle for arm I obtained using five masers
is $i=-13.0^\circ\pm2.9^\circ$, which is much more accurate than
the estimate given in the first line of Table~\ref{t1}.

Using the values of $R_k$ from Table~\ref{t2}, it is easy to
estimate the values of the wavelength $\lambda$ and radial phase
of the Sun in spiral wave $\chi_\odot$ (at $\theta=0^\circ$):
$\lambda=R_{III}-R_{II}=3.1$~kpc and
$\chi_\odot=2\pi(R_{II}-R_0)/\lambda=-140^\circ$.

From the analysis of kinematics of 44 masers done by~\cite{bb12},
the following parameters were found:
 $\lambda=2.2^{+0.4}_{-0.1}$ kpc,
 $i=-5^{+0.2^\circ}_{-0.9^\circ}$ and
 $\chi_\odot=-147^{+3^\circ}_{-17^\circ}$ for two-arm model ($m=2$) of
spiral pattern. First, we can see a very good agreement with the
result obtained in the present study for $\chi_\odot$. Second, for
four-arm model of spiral pattern ($m=4$), the value of $i$
obtained for two-arm model should be simply doubled, then
$i=-10^\circ$. Taking into account that a considerable fraction of
masers belongs to arms II, III, and the Local (Orion) arm, from
which the value of $i\approx-10^\circ$ was obtained (see Table~1
and the result by~\cite{Xu13} mentioned above), we can confirm a
good agreement with the earlier kinematic results.

It is necessary to note that, in the paper by~\cite{bb12},
the wavelength $\lambda$ was determined first by periodogram
analysis of maser residual velocities, followed by calculating the pitch angle
$i$ from $\lambda$ using Eq.~(\ref{a-04}) and
assuming $m=2$. Obviously, the results of the present work are of major
interest as we determine here the pitch angle $i$ directly, without
any assumptions on the number of spiral arms. Comparing the
estimates obtained, we can conclude that different methods give
fairly consistent results.

In the paper by~\cite{bb13}, the values $i=-6.0^\circ\pm0.4^\circ$ and
$\lambda=2.6\pm0.2$~kpc were derived from kinematic analysis of
young massive spectral binaries (of spectral classes O--B2.5) for
two-arm model of spiral pattern ($m=2$). We can see that the
consistence with results obtained in the present work is achieved
after doubling the pitch angle value.

This all means that the four-arm model of spiral pattern is most
probable in our Galaxy. Note that the scheme shown in
Fig.~\ref{f2} coincides, to within small details, with
cartographical model with parameters $i=-12.8^\circ$ and
$\lambda=3.0$~kpc for $m=4$ constructed by~\cite{Valee08,Valee13}.
Our conclusion is consistent with the one made by~\cite{Efremov11}
on the basis of analysis of the large-scale Galactic distribution
of neutral, molecular, and ionized hydrogen clouds.

\subsection[]{The Local arm}\label{Orion}
Currently, the evolutionary status of the Local arm is still not
fully clear. In paper by~\cite{Xu13}, three possibilities for the
nature of the Local arm in relation to the spiral structure of the
Milky Way were discussed in detail, with an extensive
bibliography:

\begin{enumerate}
\item The Local arm could be a branch of the Perseus arm,

\item The Local arm could be part of a Carina arm,

\item The Local arm is an independent spiral arm segment.
\end{enumerate}

These authors conclude that it is necessary to have much more
masers with measured trigonometric parallaxes in order to make a
final conclusion about the nature of the Local arm.

The value of pitch angle $i=-10.1^\circ$ found by~\cite{Xu13} from
data on masers in the Local arm is in a good agreement with the
results obtained by us for the other four arms (Table~\ref{t1} and
Table~\ref{t2}). Therefore, we can only conclude that whatever the
nature of the Local arm, its formation took place under the
influence of the Galactic spiral density wave. Following the
opinion of the majority of authors, we believe that the Local arm
is a spur rather than an independent spiral arm segment.

As we have already noted, a number of masers is associated with
proto stars or low-mass stars. It is important to know whether
there is a difference in the spiral structure parameters
determined independently from massive stars and from stars of low
masses. The reason is that, for example, a part of low-mass stars
can be formed, with a certain delay, as a result of triggered star
formation process after massive star explosions. Such a process is
observed, for example, in the nearest Sco-Cen OB
Association~\citep{Preibisch99}. To answer this question, we have
taken the example of the Local arm as it has enough masers to
identify the possible effects.

Thus we have formed two samples: 26 high-mass stars and 12
low-mass stars, so the total number of objects was 38.

We have added eight stars to the list of \cite{Xu13}. Three of
them have been included into the high-mass star sample, while five
others~--  into the low-mass sample. The high-mass stars are:
X-ray binary with a massive companion Cyg~X-1 \citep{ReidX1-2011},
maser source in the Cygnus bubble region
IRAS~20143$+$3634~\citep{Ao-2004,Yamaguchi-2012}, and radio star
HW9~CepA~\citep{Dzib11} of mass $\approx$6$M_\odot$. Note that two
masers of intermediate masses EC~95 and IRAS~22198$+$6336 in Lynds
1204G from the list of \cite{Xu13} were included in the sample of
high-mass stars.

The following masers were included Into the sample of low mass
stars: SVS/NGC~1333, IRAS~16293-2422 in $\rho$~Oph, L~1448C,
S1~Oph, DoAr21~Oph, G074.03$-$01.71, and G090.21$+$02.32 from the
list of~\cite{Xu13}. We have added also five stars from Taurus to
this sample: Hubble~4, HDE~283572, T~Tau~N, V773~Tau, and
HP~Tau/G2, according to~\cite{Torres07,Torres09,V773-2012}.

  \begin{table*}
   \begin{center}
    \caption{Parameters of linear regression $\ln(R/R_0)=a\cdot\theta+b$,
found using unit weights}
   \label{t3}
   {\small
   \begin{tabular}{|l|c|c|c|r|r|}      \hline
         Local arm & $n_\star$ & $a$       &       $b$       &      $i,$ deg &  $R_k$, kpc   \\\hline
              all  & 38 & $-0.162\pm0.010$ & $0.023\pm0.001$ & $- 9.2\pm0.6$ & $8.18\pm0.31$ \\
 hight-mass stars  & 26 & $-0.206\pm0.015$ & $0.038\pm0.002$ & $-11.6\pm0.8$ & $8.31\pm0.31$ \\
 low-mass stars    & 12 & $-0.189\pm0.005$ & $0.009\pm0.001$ & $-10.7\pm0.3$ & $8.07\pm0.31$ \\
 \hline
      \end{tabular}}
     \end{center}
   \end{table*}
  \begin{table*}
   \begin{center}
    \caption{Parameters of linear regression $\ln(R/R_0)=a\cdot\theta+b$,
found using weights $1/\sigma^2_{\ln(R/R_0)}$}
   \label{t4}
   {\small
   \begin{tabular}{|l|c|c|c|r|r|}      \hline
         Local arm & $n_\star$ & $a$       &       $b$       &      $i,$ deg &  $R_k$, kpc   \\\hline
              all  & 38 & $-0.181\pm0.005$ & $0.016\pm0.001$ & $-10.2\pm0.3$ & $8.13\pm0.31$ \\
 hight-mass stars  & 26 & $-0.244\pm0.036$ & $0.029\pm0.008$ & $-13.7\pm2.0$ & $8.23\pm0.32$ \\
 low-mass stars    & 12 & $-0.222\pm0.005$ & $0.017\pm0.001$ & $-12.5\pm0.3$ & $8.15\pm0.30$ \\
 \hline
      \end{tabular}}
     \end{center}
   \end{table*}

Results of calculations are given in Table~\ref{t3} and
Table~\ref{t4}. We can see that, when using the whole sample of 38
masers, the pitch angle value $i$ is in a good agreement with the
result obtained by~\cite{Xu13}; however, $i$ changes significantly
when splitting the sample by masses. We obtained solutions with
unit weights (Table~\ref{t3}), as well as weighted according to
distance measurement errors (Table~\ref{t4}). However, there is no
special need to apply weights for masers in the Local arm, because
almost all VLBI measurements are homogeneous. Therefore, the most
interesting result of Table~\ref{t3}, obtained from high-mass
objects, is $i=-11.6\pm0.8^\circ$. This value, as compared with
the previous one $i=-10.1\pm2.7^\circ $, turned out to be more
accurate and closer to the grand design spiral pattern pitch angle
$i=-13\pm1^\circ $, which further confirms our idea about genetic
relationship of the Local arm with the Galactic spiral density
wave.

In addition, we are interested in the parameter $R_k$. As it can
be seen from these tables, there is a small shift of $\Delta
R_k\approx0.3$~kpc, although its value is below $1\sigma$ error.
Still it seems that, for a larger number of masers, the study of
this subtle effect can give interesting results.

\section{Conclusions}\label{conclusions}
To estimate the parameters of the Galactic spiral structure~--
namely, pitch angle and the number of arms~-- we used data on
Galactic masers with known trigonometric parallaxes measured by
VLBI, with an average error of less than 10\%. These masers are
associated with extremely young objects located in active
star-forming regions.

A well-known method based on the analysis of ``position angle~--
distance logarithm'' diagram has been used. Estimates of the pitch
angle $i$ obtained from four segments of different arms belonging
to the Galactic global structure are consistent with each other
and are $i=-13.7^\circ\pm1.1^\circ$.

The most interesting one is the segment of the Outer arm, because its
objects lie in a wide range of position angles $\theta$ (up to
80$^\circ$). However, it contains only three masers, so, in order
to obtain reliable estimates, we added the data on 12 very young
star clusters with distances determined by Camargo et~al. from
infrared photometry. Using the combined data set, we determined
the value of pitch angle as $i=-11.5^\circ\pm1.9^\circ$.

Note that different methods of analysis give slightly
different estimates of the pitch angle $i$, in the
range from 10 to 14~degrees. Error of calculation of
this angle is about 1~degree. In average, the value of the
pitch angle is close to $i=-13\pm1^\circ$.

A comparison of the pitch angle $i$ found in the present study with
parameters obtained from kinematic analysis of masers (radial
phase of the Sun in spiral wave $\chi_\odot$ and wavelength
$\lambda$) allowed us to conclude that the four-arm spiral pattern
model of our Galaxy is the most probable one.

We have made an attempt to find systematic differences between the
Local arm structure parameters using
stars of different masses. This revealed some slight differences
in estimates of pitch angle, but we did not find any significant
shift across the arms (parameter $R_k$), which is probably
due to a relatively small amount of data. The most
reliable pitch angle estimate seems to be the new one
$i=-11.6\pm0.8^\circ$ obtained from the sample of massive stars.

\section*{Acknowledgments}
The authors are thankful to the anonymous referee for critical
remarks which helped to improve the paper. This work was
supported by the ``Nonstationary Phenomena in Objects of the
Universe'' Program of the Presidium of the Russian Academy of
Sciences and the ``Multiwavelength Astrophysical Research'' grant
no. NSh--16245.2012.2 from the President of the Russian
Federation. The authors would like to thank Vladimir Kouprianov
for his assistance in preparing the text of the manuscript.

\end{document}